\documentclass[
aps,%
12pt,%
final,%
notitlepage,%
oneside,%
onecolumn,%
nobibnotes,%
nofootinbib,%
superscriptaddress,%
noshowpacs,%
centertags]%
{revtex4}

\begin{document}
\selectlanguage{english}





\title{Instantaneous Radio Spectra of Giant Pulses from the Crab Pulsar from Decimeter to Decameter Wavelengths}

\author{\firstname{M.~V.}~\surname{Popov}}
\affiliation{Astro Space Center, Lebedev Physical Institute, Russian Academy of Sciences, Profsoyuznaya ul. 84/32, Moscow 117997, Russia}

\author{\firstname{A.~D.}~\surname{Kuzmin}}
\affiliation{Pushchino Radio Astronomy Observatory, Astro Space
Center, Lebedev Physical Institute, Russian Academy of Sciences,
Pushchino, Moscow oblast', 142290 Russia}

\author{\firstname{O.~M.}~\surname{Ul'yanov}}
\affiliation{Institute of Radio Astronomy, Ukrainian National
Academy of Sciences, Krasnoznamennaya ul. 4, Kharkov 61002
Ukraine}

\author{\firstname{A.~A.}~\surname{Deshpande}}
\affiliation{Raman Research Institute, C.~V.~Raman Av.,
Sadashivanagar Bangalore, 560 080 India}
\affiliation{Arecibo Observatory, HC03 Box 53995, Arecibo, 00612
Puerto Rico}

\author{\firstname{A.~A.}~\surname{Ershov}}
\affiliation{Pushchino Radio Astronomy Observatory, Astro Space
Center, Lebedev Physical Institute, Russian Academy of Sciences,
Pushchino, Moscow oblast', 142290 Russia}

\author{\firstname{V.~V.}~\surname{Zakharenko}}
\affiliation{Institute of Radio Astronomy, Ukrainian National
Academy of Sciences, Krasnoznamennaya ul. 4, Kharkov 61002
Ukraine}

\author{\firstname{V.~I.}~\surname{Kondratiev}}
\affiliation{Astro Space Center, Lebedev Physical Institute,
Russian Academy of Sciences, Profsoyuznaya ul. 84/32, Moscow
117997, Russia}

\author{\firstname{S.~V.}~\surname{Kostyuk}}
\affiliation{Astro Space Center, Lebedev Physical Institute,
Russian Academy of Sciences, Profsoyuznaya ul. 84/32, Moscow
117997, Russia}

\author{\firstname{B.~Ya.}~\surname{Losovski{\u\i}}}
\affiliation{Pushchino Radio Astronomy Observatory, Astro Space
Center, Lebedev Physical Institute, Russian Academy of Sciences,
Pushchino, Moscow oblast', 142290 Russia}

\author{\firstname{V.~A.}~\surname{Soglasnov}}
\affiliation{Astro Space Center, Lebedev Physical Institute,
Russian Academy of Sciences, Profsoyuznaya ul. 84/32, Moscow
117997, Russia}

\received{December 17, 2005}
\revised{February 6, 2006}

\begin{abstract}
The results of simultaneous multifrequency observations of giant
radio pulses from the Crab pulsar, PSR~B0531$+$21, at 23, 111,
and 600~MHz are presented and analyzed. Giant pulses were
detected at a frequency as low as 23~MHz for the first time. Of
the 45 giant pulses detected at 23~MHz, 12 were identified with
counterparts observed simultaneously at 600~MHz. Of the 128 giant
pulses detected at 111~MHz, 21 were identified with counterparts
observed simultaneously at 600~MHz. The spectral indices for the
power-law frequency dependence of the giant-pulse energies are
from ${-}3.1$ to ${-}1.6$. The mean spectral index is ${-}2.7 \pm
0.1$ and is the same for both frequency combinations
(600--111~MHz and 600--23~MHz). The large scatter in the spectral
indices of the individual pulses and the large number of
unidentified giant pulses suggest that the spectra of the
individual giant pulses do not actually follow a simple power
law. The observed shapes of the giant pulses at all three
frequencies are determined by scattering on interstellar plasma
irregularities. The scatter broadening of the pulses and its
frequency dependence were determined as
$\tau_{\mathrm{sc}}=20({\nu}/{100})^{-3.5 \pm 0.1}$~ms, where the
frequency~$\nu$ is in~MHz.
\end{abstract}

\maketitle

\section{INTRODUCTION}
\label{intr:Popov_n}

The pulsar in the Crab Nebula, PSR B0531$+$21, was discovered by
Staelin and Reifenstein [1] in 1968 as a result of detection of
anomalously strong pulses. Such pulses were subsequently called
giant. The peak flux in these pulses exceeds hundreds and
thousand times the peak flux in an average pulse formed by
synchronous integration of the pulsar signal with the neutron
star rotation period, which is equal for the pulsar B0531$+$21 to
approximately 33~ms. Giant pulses (GPs) of radio emission were
detected in several pulsars: first millisecond pulsar B1937$+$21
[2,~3], millisecond pulsars J1824${-}$2452 [4], B0540${-}$69 [5],
B1957$+$20, and J0218$+$4232 [6] as well as millisecond pulsar
J1823${-}$3021A in the globular cluster NGC~6624 [7]. GPs were
detected in the usual pulsars B0031${-}$07 [8], B1112$+$50 [9],
and J1752${+}$2359 [10]. In all these pulsars GPs are observed
rather seldom, and their properties have still been poorly
studied.

The properties of GPs have been studied in more detail for the
Crab pulsar B0531$+$21 (see, e.g., a brief review of Hankins [11])
and first millisecond pulsar [12--14]. Together with a high peak
flux density, which exceeds at decimeter wavelengths a hundred
thousand janskys, GPs from the pulsar B0531$+$21 have an extremely
short duration. At 5.5~GHz Hankins et al. [15] found in this
pulsar outbursts shorter than 2~ns. Such nanopulses have a
brightness temperature of radiation that exceeds $10^{37}$~K.
Still higher brightness temperature ($5\times 10^{39}$~K) have
been obtained for GPs from the millisecond pulsar B1937$+$21
[14]. Thus, GPs suggest a highly efficient mechanism of radio
emission, and the analysis of their properties can result in
understanding the nature of this coherent mechanism.

An important characteristic of GPs is their radio spectrum. GPs
from B0531$+$21 were detected at all frequencies where they were
searched for. For the first time GPs were detected at 112~MHz [1].
Hankins [11] mentioned observing one GP at 15~GHz.

Of primary interest are simultaneous observations of GPs at
several frequencies, when one can obtain instantaneous spectra of
individual pulses. In 1999 Sallmen et al. [16] published the
results of simultaneous observations of GPs from the pulsar
B0531$+$21 at 1400~MHz (VLA) and 600~MHz (25-m radio telescope of
Green Bank Observatory). Approximately 70\% of GPs detected at
1400~MHz were also recorded at 600~MHz. The spectral indices of
such pulses are from ${-}4.9$ to ${-}2.2$. The same paper reports
the results of simultaneous observations at 1.4 and 4.9~GHz. The
spectral indices for these frequencies were from $-4$ to $0$.
Thus, though the average value of the GP spectral index was close
to that for the components of the average pulse profile
(according to Moffett [17], in the considered frequency band this
value is ${-}3.0$), but instantaneous spectra of individual GPs
varied in rather broad limits, indicating a complex character of
the GP radio spectrum, which cannot be described by a simple
power law.

In this paper we report the results of the analysis of
instantaneous spectra of GPs in the radio emission from the pulsar
B0531$+$21 that were measured in simultaneous observations at
23.23~MHz (UTR-2 radio telescope, Kharkov), 111.87~MHz (BSA radio
telescope, Pushchino), and 600.0~MHz (64-m radio telescope,
Kalyazin).

\section{OBSERVATIONS AND DATA PROCESSING}
\label{obs:Popov_n}

The observations were carried out on November 24--26, 2003,
within the framework of the international program of
multifrequency studies of GP properties in the Crab pulsar. In
addition to the radio telescopes mentioned in the Introduction,
the GP observations were conducted also on the 100\mbox{-}m radio
telescope at Effelsberg (8350~MHz), 76\mbox{-}m radio telescope
of Jodrell Bank Observatory (1400~MHz), and Westerbork Synthesis
Radio Telescope (1200~MHz). A complete analysis of the
multifrequency observations will be presented in other
publications.

\subsection{23~MHz}
\label{UTR:Popov_n}

The observations at 23~MHz were conducted on the UTR-2 radio
telescope. The radio telescope is T-shaped and consists of three
segmented antennas: southern (S), northern (N), and western~(W),
corresponding to the directions from the UTR-2 phase center. The
UTR-2 structure uses broadband Nadeenko vibrators, which receive
radiation with the East--West linear polarization. The total
number of the UTR-2 vibrators is 2040. For more details about
UTR-2 and features of its system of amplifiers see [18,~19].

To solve our problem we have used a dual-channel reception
system. The signal summed from the S and N (S$+$N) antennas was
fed to the first channel of the receiver, while the second
channel was connected to the output of the western antenna (W).
This usage of the antennas allowed us to have an additional
criterion for the GP detection by comparison of the output signals
from the two channels. We chose identical receiving frequencies
for both channels; this allowed us to form the sum and difference
beams for the (S$+$N) antenna and W antenna.

We have used the dual-channel Portable Pulsar Receiver (PPR),
which was designed in the Laboratory of Raman Research Institute
(Bangalore, India). The receiver has an internal rubidium
standard and frequency synthesizer, due to which both channels
operate synchronously. The receiver implements digital
heterodyning of the intermediate frequency signals and four-level
(two-bit) quantizing of the output signal with a time step of
325~ns. A detailed description of this receiver can be found on
the Web site of Raman Research Institute [20].

The system noise temperature is determined by the Galactic
background radiation in the main lobe of the antenna beam, by the
radio emission of the Crab Nebula itself, and by the self-noise
of the receiving system. According to Roger et al. [21], in the
23--24~MHz band the brightness temperature of the Galactic
background toward the Crab Nebula is 40,000~K. The noise of the
UTR-2 radio telescope adds 4000\mbox{-}5000~K. The contribution
of the Crab Nebula to the noise temperature in this frequency
band is about 3000~Jy. The nonthermal pointlike radio source
(pulsar) contributes another 1000~Jy [22].

It is rather difficult to make an accurate calibration and
estimation of the UTR-2 sensitivity toward the Crab pulsar. For
the 23--24~MHz range and used UTR-2 configuration the Crab Nebula
is a pointlike object. Therefore, when observing in the decameter
range, we see fast scintillations on irregularities of the
interplanetary plasma and ionosphere. We must also take into
account the change in the radio telescope effective area due to
tracking of the pulsar.

When estimating the actual UTR-2 sensitivity, we compared in the
calibration mode the radio telescope response toward the pulsar
with the response from the reference noise generator placed at
the phase center of the radio telescope. A similar comparison was
done in additional observations of the Galactic background near
the Crab Nebula. The main beam of the radio telescope was shifted
from the Crab Nebula by approximately 2$^{\circ}$ in four
directions. To remove the effect of scintillations on the
sensitivity estimate and to determine the background level
correctly, we smoothed the data obtained during 2~h of the
observations.

This approach allowed us to estimate more correctly the brightness
temperature of the Galactic background near the directions toward
the pulsar; according to our estimates, it is $50,000\pm 4000$~K
at the frequencies 23\mbox{-}24~MHz. With the additional noise
from the Crab Nebula and pointlike source, we obtained an estimate
of the noise toward the pulsar recalculated to flux density
units, approximately 9000~Jy. With signal averaging over the
reception band (1.5~MHz) and over time (0.26624~s) we obtained an
estimate of the sensitivity, i.e., rms deviation in flux density
units, 15~Jy in the mode of registration of individual GPs. We
consider that the inaccuracy of this estimate, in view of the
errors of calibration and influence of scintillations, does not
exceed 30\%.

The effect of radiowave dispersion was removed by the method of
coherent dedispersion [23]. In the signal restoration we used the
dispersion measure 56.757~pc/cm$^3$ taken from the Jodrell Bank
Crab Pulsar Monthly Ephemeris [24]. At the frequency $f_0 =
23$~MHz in the used frequency band ${\Delta f = 1.538}$~MHz the
dispersive pulse broadening in interstellar plasma for the pulsar
dispersion measure 56.757~pc/cm$^3$ is ${\Delta t = 52.5}$~s.
When restoring the signal by coherent dedispersion on the time
interval $T$ the data segment with the duration $\Delta t$ will
be lost at the beginning or end of the total time interval $T$,
depending on whether dedispersion is applied to the lower or
upper end of the frequency band $\Delta f$. Therefore, the time
interval  $T$ on which dedispersion is implemented must
appreciably (at least by a factor of two) exceed $\Delta t$. We
performed dedispersion in ten passes, each time compensating
one-tenth of the total dispersion measure. In such an approach the
time interval $T$ was 10.9052~s, and the time $\Delta t$ for DM =
5.6757~pc/cm$^3$ was accordingly 5.253~s. The data file for the
Fourier transform had a size of 33\,554\,432 elements. Since the
expected magnitude of pulse scattering on interstellar plasma
irregularities was several seconds, we analyzed the retrieved
record of the signal with averaging over a time interval of
0.26624~s. Before the detection of the first obvious GP, the
retrieved signal was examined by eye; then we applied an automatic
procedure of isolation of suspicious events based on the
calculation of the convolution of the restored signal with a
template repeating the shape of the detected pulse. The formal
search criterion was established at a level of $5\,\sigma$ for
the values of the computed convolution. The subsequent analysis
showed that using the criterion of detection at a level of
$5\,\sigma$ yields the number of false GPs less than 10\%. This
analysis was conducted by comparing the number of events detected
in the signal that was retrieved at the final value of the
dispersion measure (56.757~pc/cm$^3$) with the number of events
detected in the signal at an intermediate stage of recovery, at
the value of the dispersion measure 45.4~pc/cm$^3$.

The time alignment of the observations at 23~MHz was provided by
injecting to the first channel of the receiver time markers bound
to the time service of the observatory. The clock rate of the
time service was checked before each observational session with
GPS signals. The marker repetition period was 10~s, the marker
duration was 20~ms, and for each whole minute an expanded marker
with a duration of 100~ms was formed. The formal accuracy of the
time alignment of the recorded data was about 0.1~ms. The
effective GP width was 4.0~seconds. The actual accuracy of the
time alignment of the detected GPs is determined by the duration
of the leading edge of these pulses and makes about 1~s.

\subsection{111~MHz}
\label{BSA:Popov_n}

The observations were carried out on the BSA transit radio
telescope of Pushchino Radio Astronomy Observatory (Astro Space
Center, Lebedev Institute of Physics) with the effective area at
the zenith of about $15\,000$~m$^2$. The flux sensitivity of the
system as measured on discrete sources and pulsars with the known
flux was $\sigma_S \approx 200$~mJy/MHz~s (toward a sky area with
the brightness temperature 1000~K). One linear polarization was
received. A 64-channel receiver with the channel bandwidth
$\Delta f = 20$~kHz was utilized. The total frequency band was
1.28~MHz. The frequency of the first (highest frequency) channel
was 111.87~MHz. The sampling interval was 8.2~ms. The time
constant $\tau = 10$~ms. Individual pulses were recorded.

The processing of the observations was performed in automatic
mode. The records were cleaned of interference, and interchannel
compensation for the dispersion delay was applied. Pulses
exceeding noise at a $5\,\sigma$ level were selected; their
amplitude, phase, and pulse width were determined.

To confirm the pulse belonging to the pulsar, we processed the
observational data with splitting into groups of different
frequency channels; processing showed the presence of a pulse in
each group and the expected dispersion delay. Another test was
processing with varied dispersion measure. As expected for the
pulsar, the pulse magnitude and signal-to-noise ratio are maximum
at processing with the nominal dispersion measure $\textrm{DM} =
56.757$~pc/cm$^3$.

The flux density was calibrated against the observations of the
pulsar B1919$+$21. For these two pulsars all the parameters of the
BSA beam pointing are identical. We have used 12 sessions of
observations of the pulsar B1919$+$21. The 111-MHz flux density
of the pulsar B1919$+$21 was adopted to be 1.55~Jy.

At 111~MHz the time alignment of the observations was done with
the rubidium standard of the observatory time service controlled
by TV signals of the State Time and Frequency Standard. The timing
accuracy was better than 100~$\mu$s. The measured effective pulse
width due to scattering is 25~ms. The accuracy of the pulse
arrival time at this frequency is also limited by the effect of
scattering and is 2~ms.

\subsection{600~MHz}
\label{TNA-1500:Popov_n}

The observations at 600~MHz were conducted on the TNA-1500 radio
telescope in Kalyazin. The radio emission was received in two
channels with left-hand (LCP) and right-hand (RCP) circular
polarizations. In each polarization channel two 4-MHz frequency
bands (upper and lower side bands) were recorded. The central
frequency is 600.0~MHz. The time resolution was 250~ns. The data
were recorded on video cassettes in a two-bit binary code by
means of the S2 recording system. The data were played back
off-line after the observations in Astro Space Center of the
Lebedev Institute of Physics on a specialized S2-RDR system,
which had been designed in Astrophysical Data Processing
Department. The equivalent system temperature is 1300~Jy in view
of the contribution of the radio emission from the Crab Nebula
[25]. To improve the sensitivity of the GP search, we averaged
the signal over 256 points (32~$\mu$s), because at 600~MHz the
temporal broadening of pulses due to radiowave scattering on
interstellar plasma irregularities is about 50~$\mu$s. With this
averaging the fluctuation sensitivity was 110~Jy (at a $\sigma$
level). In the GP search we combined in a certain way the signals
in four recorded channels, so the resulting minimum detectable
GP peak flux was 420~Jy. The technique of processing and criteria
of GP detection were described in our previous publication [26].

The time alignment of the 600-MHz observations was done with a
hydrogen standard of the observatory time service controlled by TV
signals of the State Time and Frequency Standard and by GPS
signals. The timing accuracy was 0.1~$\mu$s. At this frequency
the accuracy of the pulse time of arrival is limited by the
effect of scattering and is 10~$\mu$s. The average GP profile at
600~MHz with a time resolution of 4~$\mu$s is presented in the
figure.

\section{RESULTS}
\label{res:Popov_n}

\begin{table*}[t!]
\setcaptionwidth{\linewidth}  
\setcaptionmargin{0mm}
\caption{Observational data}
\begin{tabular}{l|c|c|c}
\hline
Frequency, MHz                &600.0      &111    &23.23\\
Time of the observations, h   &36         &0.25   &12\\
Number of detected GPs        &40\,200      &128    &45\\
Number of identified GPs      &           & 21    &12\\
$\tau_{sc}$, ms               &$0.043 \pm 0.005$       &$15 \pm 3$     &$3\,000 \pm 1\,000$\\
$S_{\mathrm{min}}^{\mathrm{peak}}$, Jy &\phantom{11\,}420&\phantom{1}60&\phantom{1}75\\
$S_{\mathrm{max}}^{\mathrm{peak}}$, Jy &34\,000       & 700   &150\\
$E_{\mathrm{max}}$, Jy~ms     &2\,000      &17\,000  &600\,000\\
\hline
\end{tabular}
\end{table*}

The table lists main observational parameters: total
observational time at each frequency, number of detected GPs,
number of GPs at 111 and 23~MHz identified with counterparts at
600~MHz, pulse scatter-broadening $\tau_{\mathrm{sc}}$, minimum
peak flux density of detectable GPs
$S_{\mathrm{min}}^{\mathrm{peak}}$, maximum peak flux density of
observed GPs $S_{\mathrm{max}}^{\mathrm{peak}}$, radiation energy
of the strongest GP $E_{\mathrm{max}}$.

Three observational sessions were carried out. At 23~MHz the
total duration of the sessions was about 1\,300\,000 pulsar
periods. During this time 45 GPs (1 GP per approximately 30\,000
pulsar periods) with peak flux densities exceeding 75~Jy, or with
energies above 300\,000~Jy~{ms}, were recorded. No GPs were
detected at 23~MHz during the time common with the 111-MHz
observations.

In three observational sessions (about 4~min each) at 111~MHz,
which included about 20\,000 pulsar periods 128 GPs (approximately
1 GP per 160 pulsar periods) with peak flux densities exceeding
60~Jy, or with energies above 2000~Jy~ms, were recorded. The energy
of the strongest GP exceeds the energy of the average pulsar
pulse by a factor of approximately 40--50.

The total duration of the observations in Kalyazin was 36 hours;
40\,200 GPs (1 GP per approximately 100 pulsar periods) with peak
flux density exceeding 420~Jy, or with energies above 25~Jy~{ms},
were recorded; 518 of them were observed in time common with the
111-MHz observations.

The smaller number of GPs detected at low frequencies is due to a
decrease in the peak flux density of the observed GPs because of
pulse scatter-smearing and increase in the brightness temperature
of the Galactic background. Therefore, only the most intense GPs
were detectable.

The figure shows examples of individual GPs at each frequency
(left) and average GP profiles (right). In this figure the time
resolution is 0.26624~s, 8~ms, and 4~$\mu$s for 23, 111, and
600~MHz respectively. The average GP profile at 23~MHz was
obtained by averaging ten strongest pulses, which were
time-aligned by the maximum of the convolution used in the GP
search at this frequency. At 600~MHz the average GP profile was
obtained by averaging all strong pulses with peak flux densities
exceeding 1000~Jy. With such averaging individual pulses were
aligned by a point at the leading edge for which the maximum
deviation from the average value was less than 30\% of the maximum
amplitude of this pulse on the record averaged over a time
interval of 4~$\mu$s. The shape of the average pulse in all three
ranges is produced by scattering on interstellar plasma
irregularities.

GPs were identified at different frequencies by coincidence of
pulse arrival instants reduced for the dispersive delay to an
infinitely high frequency with the dispersion measure
56.757~pc/cm$^3$. At 600 and 111~MHz we have identified 21
simultaneous GPs. The spectral indices in the frequency dependence
of the energies of these GPs are from $-3.1$ to $-1.6$. The mean
spectral index is ${{-}2.7\pm 0.1}$. For pulses observed at the
detection limit the spectral index for these frequencies is
$-2.6$. Since the majority of GPs detected at 600 and 111~MHz
have remained unidentified, we can conclude that 95\% of pulses
detected at 600~MHz have spectral indices with respect to 111~MHz
that are less steep than $-2.6$, while 85\% of GPs detected at
111~MHz, have spectral indices steeper than $-2.6$.

\begin{figure*}[p!]
\vskip -25mm
\hskip -20mm\includegraphics[scale=0.85]{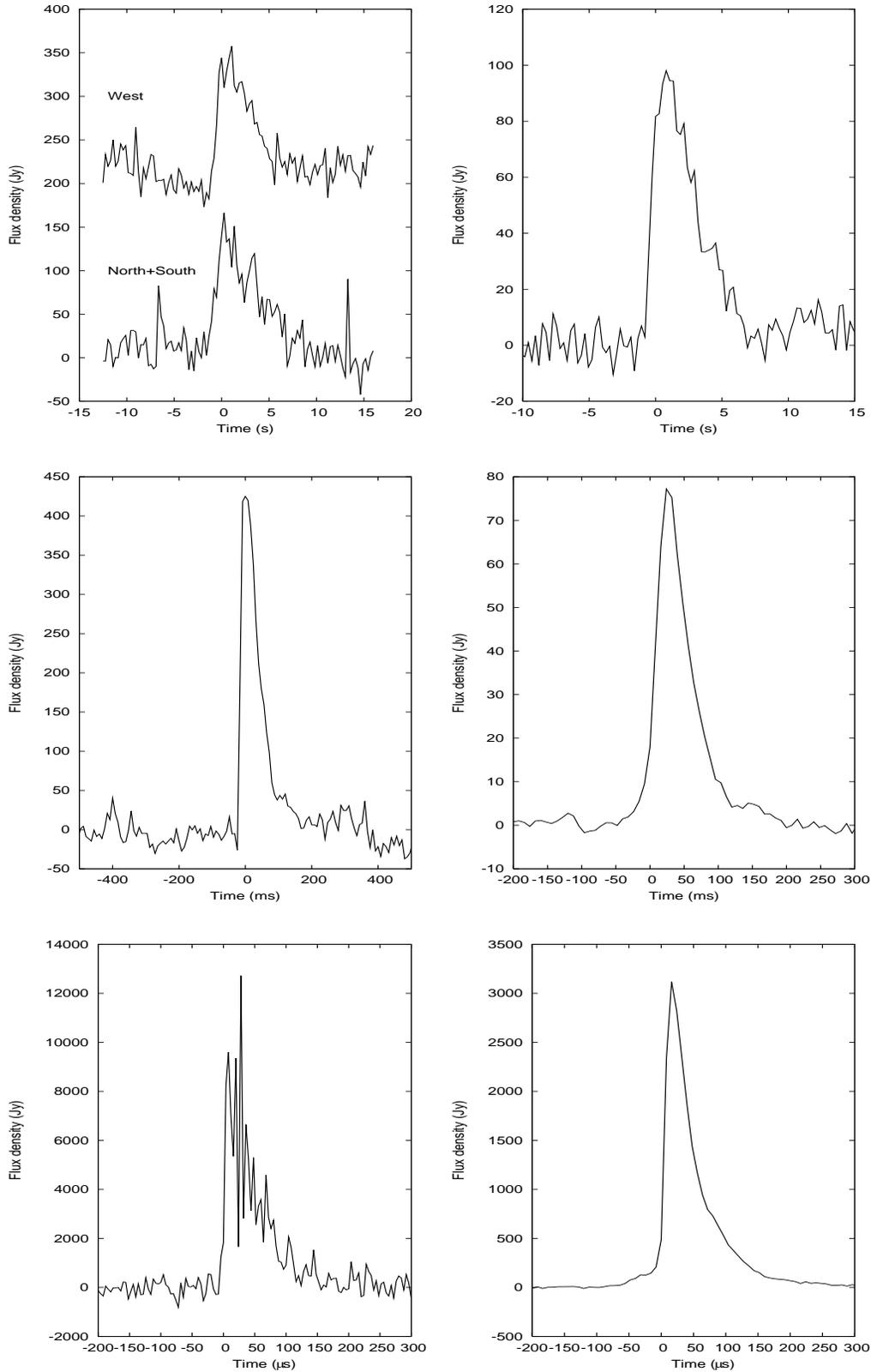}
\setcaptionmargin{5mm}
\onelinecaptionsfalse  
\caption{Upper left panel: a strong GP recorded at 23~MHz in two
channels of the UTR-2 radio telescope. Short periodic splashes in
the lower curve reflect 10-s markers injected into the receiving
channel for time alignment. Upper right panel: average profile of
GPs at 23~MHz obtained by averaging ten strongest pulses in the
channel connected to the western arm of the UTR-2 antenna. Middle
and lower panels: examples of an individual GP (left) and average
GP profile (right) at 111 and 600~MHz respectively. \hfill}
\end{figure*}

The most interesting result of our observations is the detection
of GPs at 23~MHz. To identify them with counterparts detected at
600~MHz we have supposed that GPs at 23~MHz that coincide with
GPs at 600~MHz within 1-s interval are simultaneous. To reduce the
number of spurious identifications with GPs at 600~MHz, which are
detected approximately every 3~s, we identified events at 23~MHz
only with strong pulses at 600~MHz, which exceeded the
$10\,\sigma$ level (1100~Jy). For chance events only two
identifications were expected. We identified 12 simultaneous
events. The spectral indices in the frequency dependence of the
energies of these identified GPs are from $-3.1$ to $-2.5$. The
mean spectral index is $-2.7\pm 0.1$.

This value of the spectral index should be treated with caution,
because it is subject to the selection effect. We compared at both
frequencies a small number of pulses that exceeded the adopted
detection limit. For pulses whose energies matched the detection
limits at both frequencies the spectral index is $-2.6$, while
pulses with spectral indices that appreciably differ from this
value are unobservable at one or another frequency. Since during
12~h of simultaneous observations at 600~MHz we recorded about
1500 GPs with peak flux densities above $10\,\sigma$ and only 12
pulses were identified with the events at 23~MHz, the
overwhelming majority of GPs detected at 600~MHz have spectral
indices less steep than $-2.6$. On the other hand, we can assert
on the same basis that a noticeable fraction of GPs detected at
23~MHz (about two thirds) have spectral indices steeper than
$-2.6$.

The detection of giant pulses at 23~MHz has enabled us to expand
the range for the measurement of the frequency dependence of
pulse scatter-broadening to decameter waves. GPs are observed as
very infrequent isolated pulses of a large intensity, which by
far exceeds the mean level of the pulsar radio emission. This
distinguishes GPs from a regular sequence of usual pulses,
eliminates the effect of subsequent pulses, and enables
measurements at low frequencies, where the pulse
scatter-broadening exceeds the pulsar period.

The magnitude of the pulse scatter-broadening was determined by a
comparison of the observed average GP profile with a
model-scattered template representing the initial pulsar pulse.
As a template a Gaussian pulse was taken. The template pulse was
model-scattered by a convolution with a truncated exponent
describing scattering on a thin screen. The observed pulsar pulse
was least-square approximated with the obtained model-scattered
pulse. The magnitude of scatter-broadening $\tau_{\mathrm{sc}}$,
amplitude, width, and time delay of the template pulse were the
sought parameters.

At 111~MHz the average GP profile was obtained from observations
with a higher time resolution done on the days adjacent to the
simultaneous observations. The table lists the results of the
measurement of pulse scatter-broadening. A least squares fit
yields the frequency dependence ${\tau_{\mathrm{sc}} = 20
(\nu/100)^{-3.5\pm 0.1}}$ (where $\tau_{\mathrm{sc}}$ is in ms,
and $\nu$ is in MHz), which matches well broader band
measurements of the frequency dependence of scattering for this
pulsar [27]: $\tau_{\mathrm{sc}} = 22 (\nu/100)^{-3.8 \pm 0.2}$.

\section{CONCLUSION}

We have performed simultaneous observations of giant pulses in
the Crab pulsar (PSR B0531$+$21) radio emission at 600, 111, and
23~MHz. We have detected for the first time giant pulses at the
lowest frequency 23~MHz. Thus, the mechanism of generation of
giant pulses extends to the decameter frequency band. The mean
spectral index $\alpha$ of the frequency dependence for the
energy of identified pulses is identical in the 600--111 and
600--23~MHz ranges and is equal to $-2.7$. A considerable range
of variations in the GP spectral indices and a large number of
unidentified pulses suggest that the representation of
instantaneous spectra of giant radio pulses of the pulsar
B0531$+$21 by a simple power law does not describe the true shape
of the radio spectrum of these pulses, which may have a complex
form. We have measured pulse scatter-broadening and its frequency
dependence: $\tau_{\mathrm{sc}} = 20 (\nu/100)^{-3.5 \pm 0.1}$.
The exponent of this power law differs from the values predicted
either by the Kolmogorov model for the spectrum of irregularities
of scattering plasma ($-4.4$) or by the Gaussian model ($-4.0$).
We attribute this fact to a considerable contribution of the Crab
Nebula itself to scattering of the pulsar radio emission.

\section*{ACKNOWLEDGMENTS}

The authors thank K.G.~Belousov and A.V.~Chibisov for maintenance
of the S2\mbox{-}RDR data playback system, to Yu.P.~Ilyasov and
V.V.~Oreshko for their help with the observations on the 64-m
Kalyazin radio telescope, to V.V.~Ivanova and K.A.~Lapaev for
maintenance of the observations on the BSA radio telescope, and
to A.D.~Khristenko for the help with the observations on the
UTR-2 radio telescope. This work was supported by the Russian
Foundation for Basic Research (project codes 04-02-16384 and
05-02-16415) and Program of the Presidium of the Russian Academy
of Sciences on Formation and Evolution of Stars and Galaxies. The
authors thank the Ministry of Education and Science of Ukraine,
which provided financial support of this work (contracts 729-2001
and F8/343-2004) and to Raman Research Institute (India) for
manufacturing the PPR recorder.

\end{document}